\documentclass[journal=apchd5]{achemso}
\usepackage{siunitx} 
\usepackage{tikz} 
\usetikzlibrary{decorations.pathreplacing,calligraphy,arrows.meta,positioning}

\usepackage[acronym]{glossaries}
\usepackage{xspace}
\newacronym{pso}{PSO}{Particle Swarm Optimization}
\newacronym{bo}{BO}{Bayesian Optimization}
\newacronym{bopt}{BO}{Bayesian Optimizer}
\newacronym{fl}{FL}{Fuzzy Logic}
\newacronym{fpso}{FPSO}{Fuzzy Particle Swarm Optimization}
\newacronym{afpso}{AFPSO}{Adaptive Fuzzy Particle Swarm Optimization}
\newacronym{fie}{FIE}{Fuzzy Inference Engine}
\newacronym{vt}{VT}{Viability Threshold}
\newacronym{qpso}{QPSO}{Quantum Particle Swarm Optimization}
\newacronym{cpso}{CPSO}{Classical Particle Swarm Optimization}
\newacronym{de}{DE}{Differential Evolution}
\newacronym{ga}{GA}{Genetic Algorithm}
\newacronym{gp}{GP}{Gaussian Process}
\newacronym{gbrt}{GBRT}{Gradient Boosted Regression Trees}
\newacronym{eas}{EAs}{Evolutionary Algorithms}
\newacronym{bbo}{BBO}{Black-Box Optimization}
\newacronym{cmaes}{CMA-ES}{Covariance Matrix Adaptation Evolution Strategy}
\newacronym{mo}{MO}{meta objective}
\newacronym{ei}{EI}{expected improvement}
\newacronym{pi}{PI}{probability of improvement}
\newacronym{lcb}{LCB}{lower confidence bound}
\newacronym{nfl}{NFL}{no free lunch}
\newacronym{pca}{PCA}{principal component analysis}
\newacronym{shap}{SHAP}{Shapley additive explainations}
\newacronym{cae}{CAE}{convolutional autoencoder}
\newacronym{fdtd}{FDTD}{finite-difference time-domain}
\newacronym{dnn}{DNN}{deep neural network}
\newacronym{kpls}{KPLS}{partial least square Kriging}
\newacronym{rbf}{RBF}{radial basis function}
\newacronym{agpm}{AGPM}{annular groove phase mask}
\newacronym{rcwa}{RCWA}{Rigorous Coupled Wave Analysis}
\newacronym{pcc}{PCC}{Pearson Correlation Coefficient}

\newcommand{\pso}{\gls{pso}\xspace}

\newcommand{\agpm}{\gls{agpm}\xspace}

\newcommand{\unet}{U-Net\xspace}
\newcommand{\cae}{\gls{cae}\xspace}
\newcommand{\fdtd}{\gls{fdtd}\xspace}
\newcommand{\dnn}{\gls{dnn}\xspace}

\usepackage{xcolor}
\definecolor{NordDarkBlack}{HTML}{2E3440}     
\definecolor{NordBlack}{HTML}{3B4252}         
\definecolor{NordMediumBlack}{HTML}{434C5e}   
\definecolor{NordBrightBlack}{HTML}{4C566A}   
\definecolor{NordWhite}{HTML}{D8DEE9}         
\definecolor{NordBrighterWhite}{HTML}{E5E9F0} 
\definecolor{NordBrightestWhite}{HTML}{ECEFF4}
\definecolor{NordCyan}{HTML}{8FBCBB}          
\definecolor{NordBrightCyan}{HTML}{88C0D0}    
\definecolor{NordBlue}{HTML}{81A1C1}          
\definecolor{NordBrightBlue}{HTML}{5E81AC}    
\definecolor{NordRed}{HTML}{BF616A}           
\definecolor{NordOrange}{HTML}{D08770}        
\definecolor{NordYellow}{HTML}{EBCB8B}        
\definecolor{NordGreen}{HTML}{A3BE8C}         
\definecolor{NordMagenta}{HTML}{B48EAD}       

\graphicspath{{fig}}

\title{
Photonic Structures Optimization Using Highly Data-Efficient Deep Learning:
Application To Nanofin And Annular Groove Phase Masks}

\author{Nicolas Roy}
\email{nicolas.roy@unamur.be}
\affiliation{Namur Institute for Complex Systems, University of Namur, Rue de Bruxelles 61, B-5000 Namur, Belgium}
\alsoaffiliation{Namur Institute of Structured Matter, University of Namur, Rue de Bruxelles 61, B-5000 Namur, Belgium}
\alsoaffiliation{Department of Physics, University of Namur, Rue de Bruxelles 61, B-5000 Namur, Belgium}
\author{Lorenzo König}
\affiliation{STAR Institute, University of Liège, Allée du Six Août 19c, B-4000 Liège, Belgium}
\author{Olivier Absil}
\affiliation{STAR Institute, University of Liège, Allée du Six Août 19c, B-4000 Liège, Belgium}
\author{Charlotte Beauthier}
\affiliation{Minamo Developpement Team, Cenaero, Avenue des Frères Wright 29, B-6041 Gosselies, Belgium}
\author{Alexandre Mayer}
\alsoaffiliation{Namur Institute for Complex Systems, University of Namur, Rue de Bruxelles 61, B-5000 Namur, Belgium}
\alsoaffiliation{Namur Institute of Structured Matter, University of Namur, Rue de Bruxelles 61, B-5000 Namur, Belgium}
\alsoaffiliation{Department of Physics, University of Namur, Rue de Bruxelles 61, B-5000 Namur, Belgium}
\author{Michaël Lobet}
\affiliation{Namur Institute of Structured Matter, University of Namur, Rue de Bruxelles 61, B-5000 Namur, Belgium}
\alsoaffiliation{Department of Physics, University of Namur, Rue de Bruxelles 61, B-5000 Namur, Belgium}

\begin{document}

\maketitle

\newpage
\begin{abstract}
    Metasurfaces offer a flexible framework for the manipulation of light properties in the realm of thin film optics. Specifically, the polarization of light can be effectively controlled through the use of thin phase plates. This study aims to introduce a surrogate optimization framework for these devices. The framework is
    applied to develop two kinds of vortex phase masks (VPMs) tailored for application in astronomical high-contrast imaging.
    
    Computational intelligence techniques are exploited to optimize the geometric features of these devices. The large design space and computational limitations necessitate the use of surrogate models like partial least squares Kriging, radial basis functions, or neural networks. However, we demonstrate the inadequacy of these methods in modeling the performance of VPMs. To address the shortcomings of these methods, a data-efficient evolutionary optimization setup using a deep neural network as a highly accurate and efficient surrogate model is proposed. 
    
    The optimization process in this study employs a robust particle swarm evolutionary optimization scheme, which operates on explicit geometric parameters of the photonic device. Through this approach, optimal designs are developed for two design candidates. In the most complex case, evolutionary optimization enables optimization of the design that would otherwise be impractical (requiring too much simulations). In both cases, the surrogate model improves the reliability and efficiency of the procedure, effectively reducing the required number of simulations by up to 75\% compared to conventional optimization techniques.
\end{abstract}

\newpage
\section{Introduction}
Metasurfaces offer a flexible framework for shaping the behavior of light in the realm of thin film optics\cite{mueller2017metasurface,chen2020flat,Chen2018}. 
However, their utilization in some applications necessitates intricate material configurations, posing a significant design challenge. 
To tackle this obstacle, inverse design methods fueled by computational intelligence (CI) have gained interest\cite{molesky2018inverse,lin22}.
Among them, evolutionary computation has the power to leverage deep learning within a versatile surrogate optimization framework. 
Its performance, however, should be evaluated on a challenging case. 
Design of metasurfaces such as vortex phase masks—specifically, all-dielectric phase plates\cite{arbabi2015dielectric,Bomzon02spatrot}—tailored for use in coronagraphy is certainly a relevant choice.

Coronagraphy is a powerful technique for imaging exoplanets. It enables the detection of faint planetary signals in a star's surrounding regions.
One promising coronagraphic implementation is the \agpm, which employs a focal-plane phase mask comprised of a circular subwavelength dielectric grating\cite{Bomzon02spatrot,kirill20metaphotonics,Mawet05,mawet2005subwavelength}.
The grating acts as a spatially variant half-wave plate, which creates a helical phase ramp (i.e., an optical vortex) on the optical axis of the telescope, ending-up creating a dark region in the field of view.

The performance of \glspl{agpm} has traditionally been analyzed using the \Gls{rcwa} method\cite{Moharam81}.
However, the validity of this infinite, one-dimensional grating model reaches its limits as focus shifts from the outskirts to the center of the device\cite{Konig22}.
Therefore, the use of 3D electromagnetic solvers has become necessary for modeling of the center of \Glspl{agpm}.
These new tools, by allowing more freedom in the \agpm design, provide a challenging benchmark for our framework and new ways to improve the \Glspl{agpm}.

Two approaches are devised for designing the center of the phase mask, as explained below.
Each one leads to photonic devices that are difficult to optimize due to the complex interplay between numerous mask parameters and the optical system. 
The figure of merit of the optimization is the simulated efficiency of the devices in producing the optical vortex. 
Both designs present large design spaces that require a tremendous number of 
simulations to be properly sampled: up to a trillion simulations are required to try only two values for each of the $38$ design parameters in one of the two approaches.
Therefore, we need an optimization procedure that is efficient in terms of the number of simulations required while exploring the vast search space.

In this work, we combine a \pso global optimization algorithm with a \unet surrogate model in order to optimize the mask parameters. 
This combination aims to achieve an automatic and efficient exploration of the phase mask design space. 
On one hand, \unet, a deep learning architecture borrowed from image segmentation\cite{ronneberger2015u}, is effective in providing quick (measured in milliseconds) and accurate predictions of field distributions based on the structure topology\cite{Mingkun22Wavy}.
On the other hand, \pso is a well-known global optimization heuristic algorithm.
Importantly, the methodology proposed here, while being applied to coronagraphy, is general enough to optimize any complex photonic device defined by 10 to 100 parameters with efficient exploitation of 200 to 1000 simulations. The code\footnote{https://github.com/Kaeryv/Keever} and the data\footnote{https://github.com/Kaeryv/ACSPhot23Suppl} required to reproduce our numerical experiments are available on Github.

The present work is organized as follows. 
Sections \ref{sec:eo} and \ref{sec:unet} describe the proposed U-Net surrogate modeling methodology for optimization.
Sections \ref{sec:physics} and \ref{sec:fdtd} provide an overview of the vortex phase mask coronagraph parameters and simulations. 
Section \ref{sec:sbo} describes the model enrichment process, which consists in the interaction between the optimizer, the simulation and the U-Net.
Section \ref{sec:regression} compares the implemented surrogate accuracy and efficiency with other existing models such as dense neural networks, radial basis functions (RBF) and partial least square Kriging (KPLS).
Section \ref{sec:dataconv} investigates the influence of the number of simulations used in the training of the surrogate model in order to identify a minimal dataset size.
Section \ref{sec:outcome} follows with the \Gls{agpm} designs produced by the proposed process.
Finally, Section \ref{sec:conclude} concludes the paper and discusses future research directions.

\section{Methods}
\subsection{Evolutionary optimization approach to metasurface design}
\label{sec:eo}
Global optimizers such as \pso\cite{eberhart1995particle} or \gls{ga} \cite{holland1973genetic} allows to find a solution to problems 
that otherwise would require an extremely high number of simulations\cite{Mayer22uwba}. 
For the design of metasurfaces, this can result in unreasonably high computational needs, as they most often require complex solvers such as \gls{fdtd}.
However, global optimizers still require thousands of figure of merit evaluations in order to find an optimal combination for several tens of variables.
Using a simulation result as figure of merit implies that each evaluation lead to a full numerically expensive simulation.
When using surrogate models\cite{SMT2019,wang2019,forrester2009}, the amount of simulations required is significantly reduced by partially relying on the quick predictions it provides for evaluation.
Surrogate models also have other interesting properties\cite{yew06curse}: a simpler model of a complex simulation will offer a smoother performance metric to the optimizer.

We chose this global evolutionary optimization approach over more popular inverse design methods\cite{Mingkun22Wavy, ACSNeuralFourier,Hammond22AdjointMeep} due to several reasons. 
First, the surrogate optimization approach provides an interesting compromise between the fast convergence of adjoint solvers\cite{Hammond22AdjointMeep} and the global exploration of the design space allowed by global optimizers. 
Second, a metasurface design consists in a tight geometrical description of the device using bounded parameters contrary to inverse design techniques that use a freeform density-based description\cite{Hammond22AdjointMeep}. Finally, the surrogate optimization scheme is straightforward to implement, consistent across different devices and it enables massively parallel workloads for the full solver. Other types of parameters such as categorical variables can easily be implemented for materials and shapes.

\subsection{The surrogate solver: U-Net}
\label{sec:unet}
Various black box methods can be substitutes (surrogate models) for simulations, including interpolation techniques such as \glspl{rbf}, Kriging\cite{rbf90,bouhlel2016improving} or other classic machine learning methods such as regression trees\cite{regtrees,xgboost}.
In recent years, \glspl{dnn} were often used as surrogate models\cite{sun2019review,kochkov2021machine}, particularly in photonics\cite{chen22wavey}.
These \Glspl{dnn} are described as universal approximators \cite{cybenko1989approximation,lu2020universal} for any continuous bounded function, making them a versatile tool used in many tasks.
In fact, \Glspl{dnn} can take many forms, referred to as architectures, to adapt to the data of the problem at hand.

\begin{figure}
  \centering
  \resizebox{0.9\columnwidth}{!}{\input{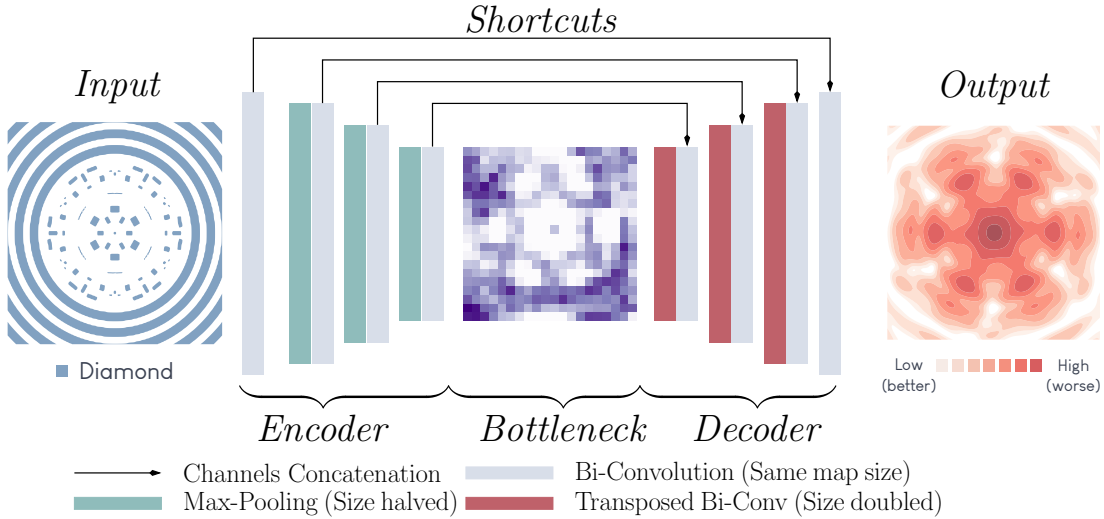}}
  \caption{\textbf{Representation of the \unet architecture.} The network is comprised of an encoding and a decoding module. These are located at each side of a bottleneck. The input of the network is a slice of the dielectric structure (left) while the output is a conform slice of the polarization leakage field (right). The shortcuts directly connecting layers along the encoder and decoder are represented by the black arrows.}
  \label{fig:unet}
\end{figure}
In this work, we will be using an architecture of \dnn called the \unet\cite{ronneberger2015u} implemented using PyTorch\cite{pytorch}.
This architecture, illustrated in Figure \ref{fig:unet}, excels in image-translation tasks such as image segmentation, where the
spatial topology of the image is preserved while the meaning of each pixel changes. 
In our case, the input is the relative dielectric permittivity and the output is a real function of the electromagnetic fields, namely the leakage field.
While the \unet is able to handle 3D data \cite{cicek16}, a 2D representation of both the structure and the polarization leakage field is numerically more efficient and sufficient to reproduce the physics of the whole system.
Therefore, characteristic 2D slices of the dielectric and polarization leakage field were used as represented in Figure \ref{fig:unet}. The slices go through a sequence of convolution operations until they reach a low dimensional representation at the center (bottleneck) of the \unet (Figure \ref{fig:unet}).
This compressed representation is then transformed back to match its original size at the output, using transposed convolution operators.

Up to this point, the \unet is very similar to a \cae \cite{Masci11}. 
However, the  \unet is completed by skip-connections (\emph{shortcuts} in Figure \ref{fig:unet}), 
allowing low-level, local features to pass through and merge with 
features emerging from the bottleneck, along the decoder. 
These shortcuts make the \unet effective in modeling the behaviour of electromagnetic fields inside metasurfaces as they link more easily the fields to the local dielectric topology.
This network architecture has been shown recently to perform well for a diffracting system\cite{chen22wavey}.

\subsection{Metasurfaces: the case of annular groove phase masks}
\label{sec:physics}
\begin{figure}[b!]
  \centering
  \includegraphics[width=0.49\textwidth,page=2]{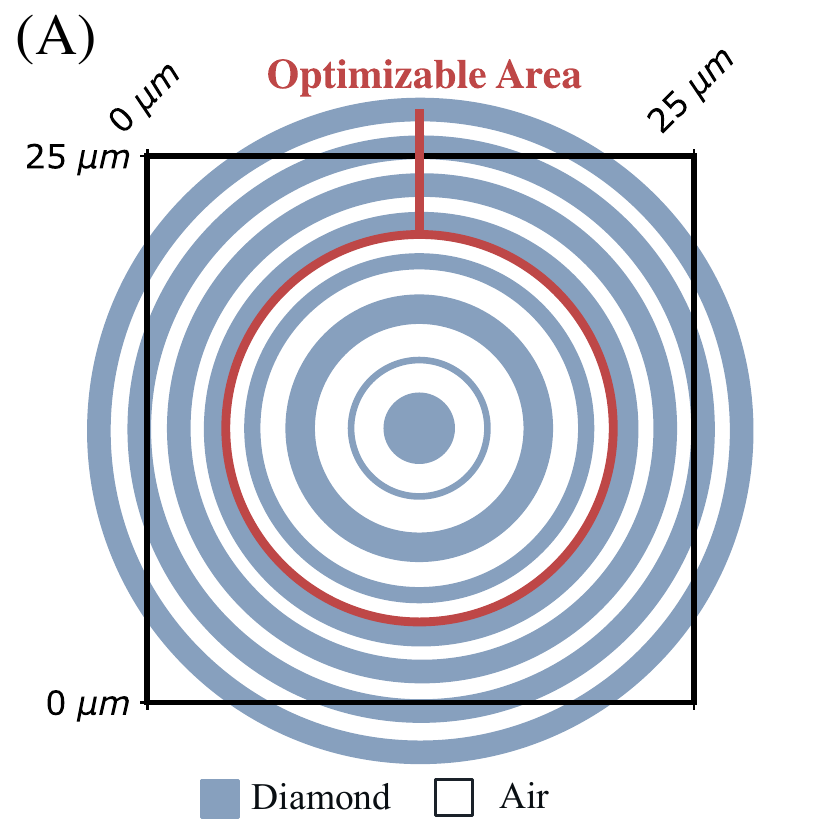}
  \includegraphics[width=0.49\textwidth,page=3]{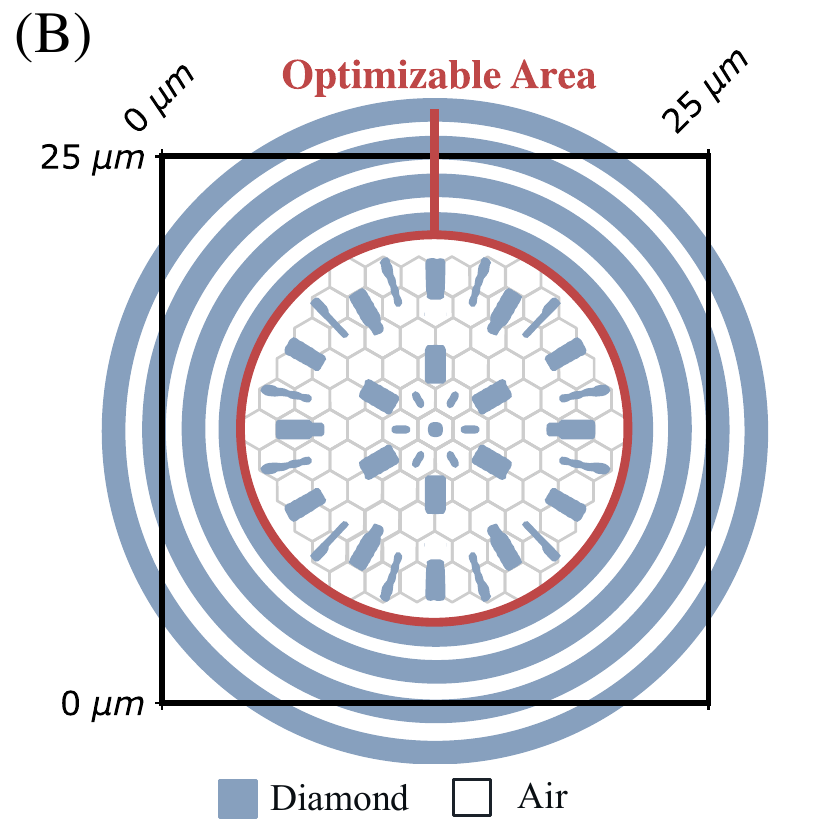}
\caption{Dielectric permittivity ($\epsilon_r$) profile of an (A) annular groove and a (B) nanofin phase mask. The \agpm pattern is maintained in the region beyond the optimizable area (red circle) and determines the depth of the full pattern. The hexagonal pixelization grid used to define the metasurface pattern is overlaid as thin grey lines in (B). } 
  \label{fig:agpm1}
\end{figure}


The optimization framework presented in this paper is applied to the design of the AGPM center.
The \gls{agpm} is one way of implementing a vector vortex phase mask capable of achieving contrasts of several orders of magnitude across a large bandwidth \cite{Mawet05,Delacroix12AppOpt}. 
Several \glspl{agpm} have been successfully installed on the world's most advanced telescopes to date\cite{Delacroix12VISIR,Mawet13NACO,Defrere14LMIRCam,Serabyn17Keck}.
The \agpm uses the artificial birefringence of subwavelength gratings etched onto a diamond substrate to create the helical phase ramp characteristic of a vortex phase mask. 
The grating parameters are tuned to provide a $\pi$ phase shift between orthogonal polarizations and create an achromatic half-wave plate. 
The characteristic helical phase ramp is obtained by spatially varying the fast axis of the half-wave plate across the mask, resulting in the circular grooves pattern of the \agpm. 

\subsection{The physics solver: Finite-difference time-domain simulations}
\label{sec:fdtd}
While the grating parameters of the \agpm are optimized using RCWA \cite{Moharam81}, which is well suited for describing infinite periodic gratings, at the center of the \agpm, the pattern is no longer periodic.
The \gls{fdtd} method \cite{Oskooi10} is used here to fully describe the behavior of the \agpm at its center. 
A circularly polarized plane wave is propagated through the \agpm. The half-wave plate character of the \agpm flips the helicity of the wave while imprinting the textbook helical phase ramp, leaving a small fraction unaffected referred to as polarization leakage. 
The polarization leakage is numerically computed as the mean intensity of the circular polarization with the same handedness as the input in a slice \SI{2.25}{\micro\metre} inside the substrate. This polarization leakage quantifies the amount of light that does not acquire the phase ramp due to the chromaticity of the design.
Effects of the curved grating lines near the AGPM center can be described accurately and an optimal size can be estimated for the central pillar\cite{Konig22}.

Here, the \agpm center is inversely designed by using a more complex metasurface structure. 
Starting from the \agpm pattern, a region with a radius of five grating periods is optimized. 
These five periods correspond to the region in which the central leakage is localized (optimizable area in Figure \ref{fig:agpm1}).
A simple pattern of concentric grooves of varying line width and position is first considered to minimize the leakage term at the center of the mask. 
Figure~\ref{fig:agpm1}(A) shows an example of a concentric groove pattern defined by its inner and outer radius for each groove. 
The design freedom is then drastically increased by using rectangular nanofins \cite{chen2020flat} placed in a hexagonal pixelization grid (Figure \ref{fig:agpm1}(B)), leading to an optimization problem with hundreds of free parameters. 
Each nanofin in the pixelization grid has 5 free parameters: 
its position (2), size (2) and its tilt angle (1). 
In principle, for 91 blocks shown in Figure \ref{fig:agpm1}(B) within the optimized region this results in 455 free parameters. 
However, for the case of the \agpm, the circular symmetry of the problem is exploited by using the symmetry of the hexagonal pixelization grid. 
Forcing the orientation of the blocks to be parallel or orthogonal to the annular grooves further reduces the number of free parameters: 
while keeping the vectorial nature of the mask, the number of independent parameters is reduced to 38. 
Figure~\ref{fig:agpm1}(B) shows an example of a nanofin pattern, including the hexagonal pixelization grid overlaid as thin grey lines. The depth of the structures is fixed throughout the mask for both patterns and was optimized for the annular grooves pattern beyond the central region considered for optimization. Such parameters are available in supporting information S3.

\newpage

\subsection{The global optimization scheme}
\label{sec:sbo}
\begin{figure}
  \includegraphics[width=1.0\textwidth]{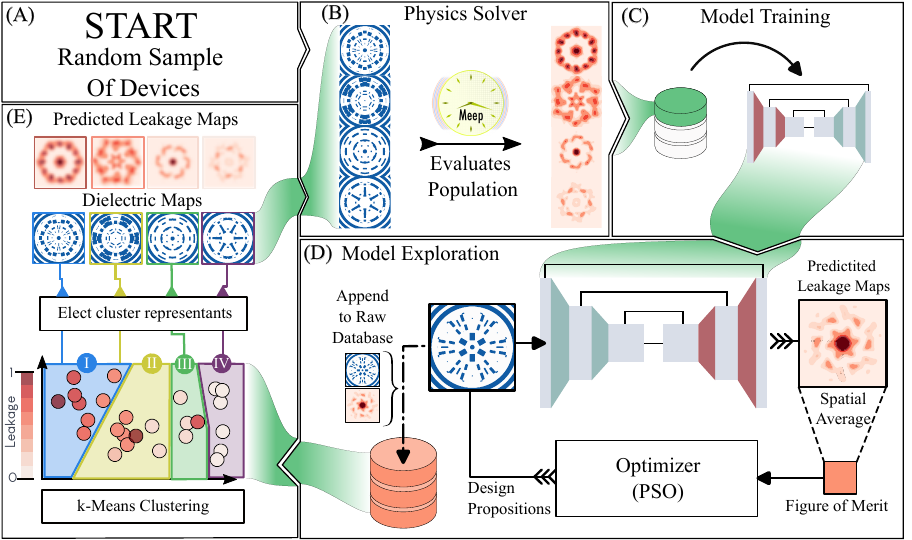}
  \caption{\textbf{Summary of the surrogate based optimization scheme} (A) To initiate the algorithm, a randomly generated selection of designs is formed. (B) The selected designs are evaluated on the expensive \fdtd simulations. (C) The new designs are appended to the dataset and the U-Net is updated. (D) A \pso algorithm searches for better designs using the \unet model. (E) The designs proposed by \pso are filtered using k-means clustering, a design is elected in each cluster.  The loop is closed by going back to step (B).}
  \label{fig:summary}
\end{figure}

In this section, a surrogate optimization process is presented. This process will iteratively feed the \unet with new designs and leakage maps computed through \gls{fdtd} to improve its accuracy.
The choice of those designs is left to a \pso algorithm \cite{fuzzy22} that will probe designs with increasing performance.
The whole model building and optimization process is illustrated in Figure \ref{fig:summary}.

The process starts by generating $N_0$ random designs (Figure \ref{fig:summary}(A)). This value is fixed following a study in an upcoming section ($N_0=50$).
These initial designs enter the main loop and are evaluated using \gls{fdtd} simulations, as described in Section \ref{sec:physics} (Figure \ref{fig:summary}(B)).
Once this first batch of simulations is performed, a first \unet is trained on the initial set of simulations (Figure \ref{fig:summary}(C)).
The \unet will then be explored by multiple ($n_{\text{pso}}$) parallel executions of the \pso algorithm (Figure \ref{fig:summary}(D)). 
This exploration leads to pairs consisting of the design that is probed by PSO and the corresponding leakage maps predicted by the \unet.
PSO being a stochastic algorithm, each instance will lead to a different path and sequence of probed designs.
These design-prediction pairs are then stored in a raw database for further analysis (Figure \ref{fig:summary}(D)).

This raw database cannot be directly used to improve the \unet as its performance is only an estimation made by the \unet itself. As such, it may include significant errors, especially at the early stages of training. 
Furthermore the dataset is large and redundant. For all these reasons, a selection process (Figure \ref{fig:summary}(E)) is required to pick just a few designs that are allowed to go back to evaluation step (B).
Thanks to this selection, only the most relevant of these designs will be associated with correct leakage maps in step (B) of Figure \ref{fig:summary} using the \gls{fdtd} solver.

Relevant designs are those that provide information about the search space (diversified) and achieve optimal optical performance.
Failing to balance exploration (attempting less promising designs) and exploitation (refining the best ones) would lead to either early convergence on a suboptimal design or failure to converge altogether.
To make the selection, the raw database is split in $k$ clusters by applying k-means \cite{hartigan1979k} ($k=4$) on the topology of the designs as illustrated in Figure \ref{fig:summary}(E).
The approach works in synergy with \Gls{pso} due to the observed tendency of particle swarms to form a sequence of design niches during the optimization process of step (C). Evidence for this statement is available in supporting information S2. 
The clusters of k-means then allow us to sample these niches as they each correspond to a local optimum for the design coined archetype.
Finally, a design is picked with a probability inversely proportional to its performance in each cluster.
More details about the selection can be found in supporting information S2.

Once filtered, the few remaining designs take the same path as the initial random sample:
they are evaluated accurately with \gls{fdtd} in \ref{fig:summary}(B), the pairs of dielectric and leakage maps are stored in the database (C). 
The \unet model then incorporates the increased dataset in \ref{fig:summary}(C). This new model is finally used in (D), like the initial one, closing on the surrogate optimization loop.

Using $n_{\text{pso}}=5$ parallel \pso instances, with one design sampled in each of $k=4$ clusters, turned out to perform well, 
leading to $k \times n_{\text{pso}}=20$ \fdtd evaluations for each model update. From 15 to 20 model updates were found to be sufficient for the optimization to converge.
These parameters were used in the optimization of nanofins (NF) and annular groove (AG) phase masks.

\section{Results}
\subsection{Quality of the polarization leakage predictions}
\label{sec:regression}
In the previous sections, the \unet was introduced as a promising architecture for surrogate modeling of complex metasurfaces.
In this section, this statement is supported by comparing the \unet approach to competing methods for the prediction of the spatially averaged polarization leakage.

\begin{figure}
  \centering
  \includegraphics[width=0.95\textwidth]{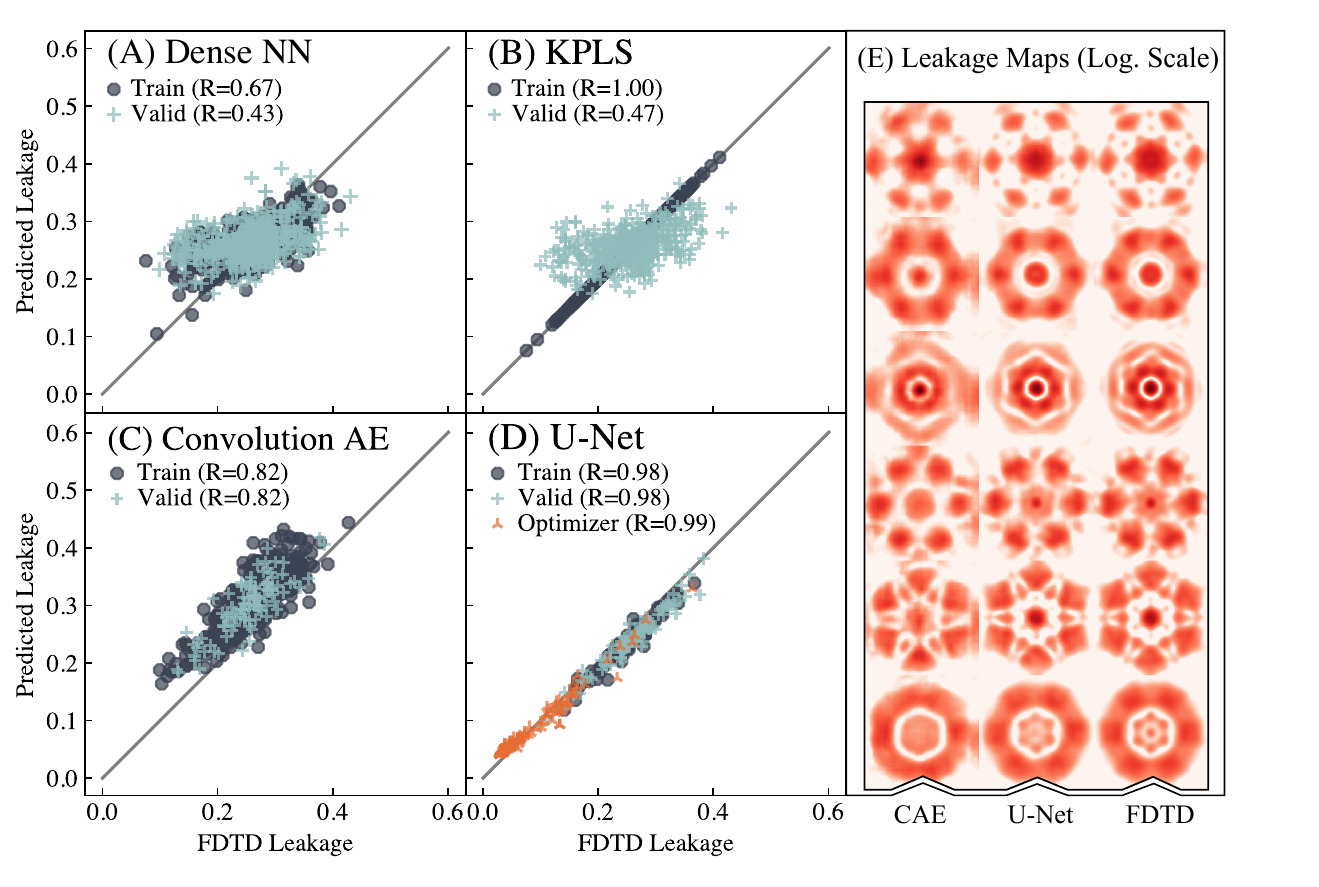}%
  \caption{\label{fig:qqplot} \textbf{Benchmark of \unet against dense neural network, KPLS and \gls{cae}.}
  (A-D) Scatter plots of the predicted versus the \Gls{fdtd} mean leakage from various methods. 
  The ground-truth leakage computed by \gls{fdtd} is found on the abscissa while the ordinates represent the corresponding predicted values.
  The results for two groups of designs are shown and correspond to the training set and the validation set. Predictions made during optimization are also shown for the \unet. The Pearson Correlation Coefficient (R) is displayed in the legend. (E) CAE, \unet and \Gls{fdtd} leakage maps are shown for six random designs.}%
\end{figure}

Figure \ref{fig:qqplot} shows, for the nanofins designs, the predicted spatial average of the polarization leakage ($\bar l$) against its ground-truth value obtained through the \gls{fdtd} method.
Results for different groups of designs are shown in Figure \ref{fig:qqplot}(A-D):
the \emph{training dataset}  used for building each model and the \emph{validation dataset} that is unknown to each model.  
The \emph{optimizer dataset}  represents all designs probed during an optimization session and is only shown in Figure \ref{fig:qqplot}(D) as red nails.
The training and validation sets were the same for all methods and contained 2500 designs each. While the training or validation set is randomly sampled, the \emph{optimizer} set 
is biased towards better designs not found in a random sampling as can be seen in Figure \ref{fig:qqplot}(D) by the lower values of $\bar l$.

The \unet showed consistent accuracy over the training, validation and optimizer datasets with a \gls{pcc}\cite{pcc} $R$ of respectively $0.98$, $0.98$ and $0.99$. The \cae reached the second best performance with a correlation coefficient of $0.82$.
This is expected as the \cae and \unet share the way they handle the inference: they match spatial distributions (i.e., maps) of the dielectric permittivity and leakage fields. 
Maps of the dielectric permittivity are directly built from the design parameters described in Section \ref{sec:physics}, while the produced leakage field maps are spatially averaged to obtain the mean leakage $\bar l$.  
The relationship between the dielectric and the leakage field spatial distributions proves to be well modeled as shown in Figure \ref{fig:qqplot}(E). 

The performance gap between \cae and \unet can be observed through the predicted leakage maps of Figure \ref{fig:qqplot}(E). 
The only architectural difference brought by the \unet is the presence of skip-connections. 
Thanks to these, the \unet predicts finer local design features accurately, which is an expected property of the \unet\cite{ronneberger2015u}. 
Some non-physical field artefacts are present in the \cae predictions while they do not appear with the \unet.

Contrasting with \cae and \unet approaches, \glspl{dnn} and \gls{kpls} regression models \cite{bouhlel2016improving,SMT2019} attempt to predict the mean leakage $\bar l$ directly starting from the geometric parameters of the dielectric structure. 
These approaches failed, barely reaching a moderate correlation ($R=0.5$) with this large training dataset. This poor result is mainly due to the complexity of the interaction between the geometric structure parameters and the mean leakage. \gls{kpls} interpolation is particularly ineffective as the mean leakage reacts abruptly to many of the geometric parameters.

Our approach yielded a robust deep learning model, capable of being repurposed for future predictions, even when applied to a similar but different devices.
Specifically, the model trained on nanofins serves as a solid foundation for making predictions in the annular groove case and vice versa. Moreover, this network is also valuable for tasks such as inverse design or SHAP (Shapley additive explainations) analysis, as introduced in the work of Lundberg et al. (2017) \cite{Lundberg17SHAP}.

The key takeaway message is the superior accuracy of the \unet, with $R\ge 0.95$.
These results strongly justify the choice of \unet as a surrogate for \gls{fdtd} simulations in photonics, particularly when characteristic slices can be extracted to reduce the complexity and size of the model.

\newpage
\subsection{Influence of the dataset size}
\label{sec:dataconv}
While the accuracy displayed by the \unet is compelling for optimization, its benefits have to be balanced against the number of simulations required to train it. 
In fact, machine learning models generally require thousands of simulations, with a great dependence on the problem at hand. It is difficult, if not impossible, to meet this expectation for numerically expensive simulations. This study aims to define the minimal number $N_0$ of simulations required to initiate a coarse \unet model that brings the validation \Gls{pcc} ($R$) above $0.5$. This particular value was selected based on observation of the field maps, and corresponds to a ``moderate correlation'' in statistics.

\begin{figure}
  \centering
  \includegraphics[width=\textwidth]{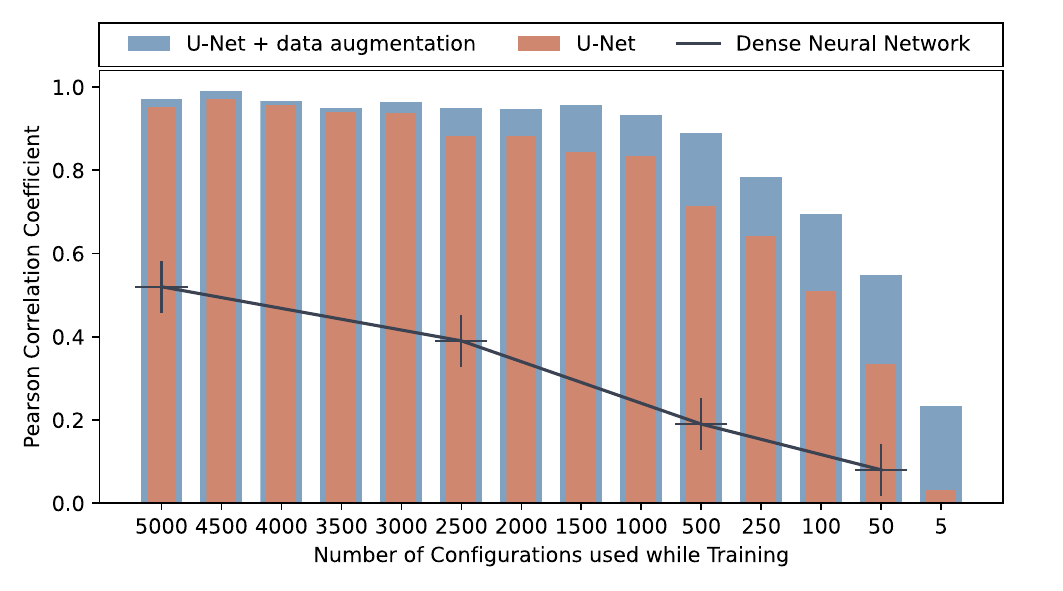}%
  \caption{\textbf{Regression quality with varying dataset size.} The \Gls{pcc} on the validation set is shown for different sizes of the training set. Five randomly seeded trainings were considered for each dataset size. The data augmented training (blue) is compared to a standard training (orange).
   The performance of the simpler dense neural network is charted in black for benchmark.}%
  \label{fig:dataconv}
\end{figure}

The efficiency of the \unet predictions is assessed in Figure \ref{fig:dataconv} where the model is trained with increasingly scarce training data. The validation set \gls{pcc} ($R$) is plotted against decreasing number of training set simulations. The training and validation sets are obtained by splitting a common source dataset of 5000 simulations from randomly chosen designs.

Figure \ref{fig:dataconv} shows that \unet maintains a $R\ge0.5$ for the validation set for as few as $N_0=100$ simulations in the training set. 
Even better results can be achieved through data augmentation.
Data augmentation\cite{shorten2019survey} helps to improve the robustness of machine learning models by exposing them to a greater variety of data, which enables them to better generalize to unseen data by coping with noise, variations, and biases. 
In our case, the data is augmented by cropping and rotating the designs and corresponding polarization leakage maps at random angles while training. 
\unet with data augmentation is shown to achieve $R\ge0.5$ with as few as $N_0=50$ simulations (Figure \ref{fig:dataconv}, on the right).
For large simulation datasets, on the other hand, data augmentation has a limited impact on the validation \gls{pcc} (Figure \ref{fig:dataconv}, on the left).
This is expected as the available data become sufficient to fully train the network. 
In short, data augmentation proves to be a key tool to handle the small dataset sizes at the beginning of an optimization process. 

\subsection{Performance analysis of vortex phase masks designs}
\label{sec:outcome}
\begin{figure}
  \includegraphics[width=0.99\textwidth,page=1]{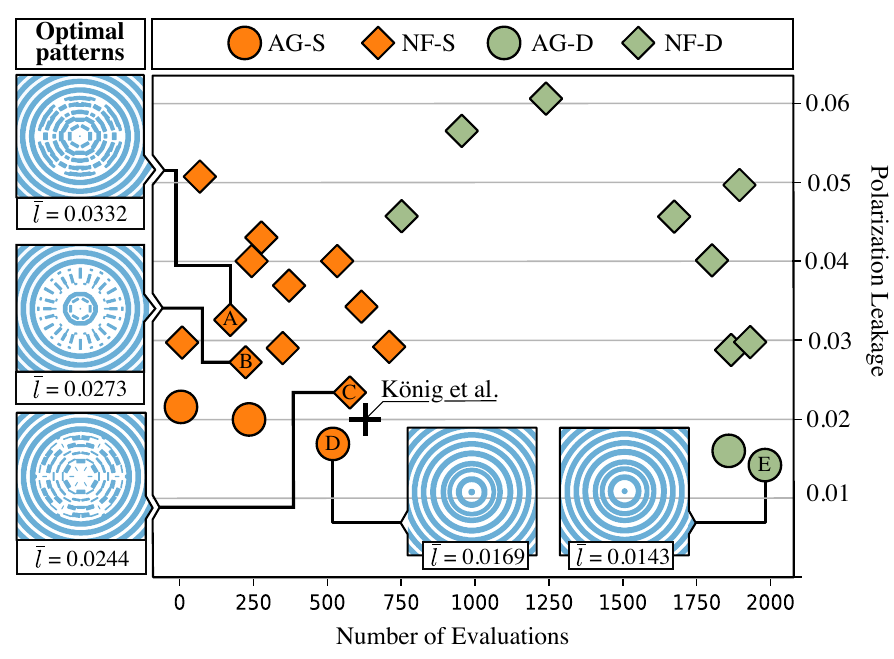}
  \caption{\textbf{Overview of the optimization performance.} 
  Optimal polarization leakage is plotted against the number of evaluations required in four optimizations schemes. 
  An overview of some designs, including the best one for both the AG-S and NF-S processes, is provided on the left. Previously published results\cite{Konig22} are plotted as \textit{König et al.}}
  \label{fig:results}
\end{figure}

Figure \ref{fig:results} compiles the optimizations of the Annular Groove (AG) and the Nanofins (NF) vortex phase mask designs for the above discussed surrogate optimization method and the direct \pso optimizer (directly using the \Gls{fdtd} solver instead of the \unet to evaluate the figure of merit $\bar l$). 
The four resulting processes are labeled as follows: \textbf{AG-D} and \textbf{AG-S} optimize the Annular Groove design with direct and surrogate optimizers respectively, while \textbf{NF-D} and \textbf{NF-S} optimize the nanofins pattern.
Each marker in Figure \ref{fig:results} corresponds to a full optimization process. 
For efficient use of computational resources, optimizers were stopped when no progress happened during 40 evaluations using simulations. 
This condition corresponds roughly to two iterations for the surrogate optimizer (AG-S,NF-S) as well as for the direct optimizer (AG-D,NF-D).

Figure \ref{fig:results} shows how efficient our surrogate optimizer approach is when compared to direct optimization. In the \textbf{NF-S} case, surrogate optimization finds equivalent or better solutions in four times less evaluations than \textbf{NF-D}. The surrogate method is also more reliable, two thirds of processes ending up with a polarization leakage below $0.04$ while the direct optimizer only has one fourth.

Another noteworthy aspect is the reduced leakage observed in annular groove designs, suggesting that the concentric rings are superior in converting right-handed circular polarization and generating the anticipated helical phase ramp compared to the nanofins.
This also appears in the inset nanofins designs, for two of them (A,C) mimic an annular groove.
This systematic approach confirms that the highest geometrically-induced anisotropy is achieved by the 1D grating, even at the center.

For the annular groove optimization, results are similar on average to the ones reported in a previous work \cite{Konig22} ($\bar l = 0.02$), where an annular groove pattern (defined by two parameters) was devised by plotting the leakage value for 25 values of both parameters, amounting to 625 evaluations. Best designs of the direct optimizer (E) managed to obtain leakage $\bar l$ of $0.0143$, which represents a 25\% improvement with respect to previous work\cite{Konig22}. 
Meanwhile, the surrogate optimizer enabled a slightly lower (better) leakage (D) of $0.017$ with 
a similar simulation budget.
The superior efficiency of the direct optimizer in this case is expected since it operates within a smaller search space, consisting of only 10 free parameters, compared to the nanofins design, which involves 38 parameters. Moreover, designs (D,E) are very practically similar, suggesting that the difference of performance originates from small variations in the design. Still, the surrogate optimizer found it with four times less evaluations.

While many produced designs tend to imitate an annular groove pattern such as in Figure \ref{fig:results}(A,C), some of the best performing ones use a mix of agglomerated fins parallel or orthogonal with the external annular groove. 
This is a sensible design as the birefringence can also be produced by a radial pattern.

\section{Conclusion}
\label{sec:conclude}

This work demonstrates the potential of global optimization methods in a new era of photonic design lead by novel machine learning and adjoint optimization methods.
Specifically, we have highlighted the role of evolutionary optimization (PSO), which enables a stable global exploration of the search space.
The staggering need for evaluations of evolutionary algorithms was alleviated thanks to a \unet surrogate solver.

An already efficient surrogate \gls{fdtd} solver (an \unet) was made even more efficient thanks to data augmentation.
This surrogate provided swift and accurate evaluations of designs for a PSO global evolutionary optimization algorithm.
The \unet worked by matching characteristic slices of the design and figure of merit instead of considering the whole simulation domain, resulting in a simpler and faster model.
The resulting surrogate optimization framework enabled  optimization of designs with four times less simulations than required by the evolutionary algorithm.
The optimization scheme was also shown to be more reliable given the lower variance in the results obtained when compared to direct optimization of the simulation.

The surrogate-based optimization was applied on two types of devices: an annular groove and a nanofins pattern. 
An optimum was identified in the former case, achieving performance that is either comparable or surpasses previous attempts. 
This optimization scheme is versatile: it could integrate other types of design parameters such as categorical parameters for choosing materials and shapes aside from geometric lengths.
Moreover, as this scheme makes no preliminary hypothesis on the underlying simulation, it can be applied to a wide range of photonic design problems. 

\bibliography{c23_main}
\section{Supporting Information}
Dielectric and leakage profiles for some of the best designs (S1). Details for the reproducibility of the surrogate optimization procedure (S2). Details for the reproducibility of the FDTD simulations (S3).
\section{Acknowledgments}
  Computational resources have been provided by the Consortium des Équipements de Calcul Intensif (CÉCI), funded by the Fonds de la Recherche Scientifique de Belgique (F.R.S.-FNRS) under Grant No. 2.5020.11 and by the Walloon Region. 
  This project has received funding from the European Research Council (ERC) under the European Union’s Horizon 2020 research and innovation programme (grant agreement No 819155).
  A.M., M.L. and O.A. are funded as Research Associate (A.M. and M.L.) and Senior Research Associate (O.A.) by the Fund for Scientific Research (F.R.S.-FNRS) of Belgium.
  The authors also thank Pr. Olivier Deparis (University of Namur) for generously dedicating their time and expertise to meticulously proofread this manuscript. Their insightful feedback and attention to detail greatly enhanced the clarity and accuracy of our work.
\end{document}